\documentclass[jkps,preprint,fleqn,showpacs,showkeys]{revtex4}
\usepackage{graphicx}
\usepackage{amssymb}
\usepackage{amsmath}
\usepackage{bm}
\begin{document}
\setcounter{page}{1}
\title[]{Light Dragging Phenomenon and Expanding Wormholes}
\author{Hristu \surname{Culetu}}
\email{hculetu@yahoo.com}
\thanks{Fax: +40-241-618372}
\affiliation{Department of Physics, Ovidius University, \\B-dul Mamaia 124, 900527 Constanta, Romania}

\date[]{Received 27 May 2010}

\begin{abstract}
 The null geodesic congruence for the Lorentzian version of Hawking's wormhole is studied in spherical Rindler coordinates. One finds that the wormhole throat, where the stress energy is mostly located, expands exponentially and that the flare - out condition is satisfied. A time reversal is equivalent to an inversion applied to the radial coordinate. Far from the throat (the light cone in Cartesian coordinates, the negative energy density acquires an expression similar to that of the Casimir energy density between two perfectly reflecting plates. We conjecture that the light propagation follows the throat of a pre - existing expanding wormhole (a light dragging phenomenon).
\end{abstract} 

\pacs{04.50.+h; 04.60 .-m; 04.90.+e}.
\keywords{Dynamic wormhole, Conformal Rindler geometry, Null geodesic expansion, Spatial inversion}.

\maketitle

\section{INTRODUCTION}

Lorentzian wormholes were often considered as topological objects (handles in spacetime that join two distant regions of the same universe or bridges linking different spacetimes) \cite{HV}. Both of them give rise to multiply - connected universes with non - trivial topology and are hypersurfaces that connect two asymptotically flat spacetimes.
 
 As Hochberg and Visser \cite{HV} have remarked that to define a wormhole, one identifies a local property that is operative near its throat through the expansion of the null geodesic congruence propagating orthogonally to a closed two - dimensional spatial hypersurface, generalizing in this way the Morris - Thorne analysis \cite{MT}. In their view, the hypersurface will be the wormhole throat provided the expansion $\Theta$ of the null congruence vanishes on it and the rate of change of the expansion along the null direction $\lambda$ is such that $d\Theta/d\lambda \geq 0$ (the flare - out condition, which is valid even for dynamic wormholes). In other words, the throat is a surface of minimal area with respect to deformations in the null direction $\lambda$; \textit{i.e.}, the throat is an anti - trapped surface.
 
 The flare - out condition at or near the throat violates the null energy condition that results directly from the Raychaudhuri equation \cite{KS}. Guendelman \textit{et al.} \cite{GKNP} showed recently that the Einstein - Rosen bridge metric \cite{ER} is not a solution to the vacuum Einstein's equation, but rather a lightlike brane source at $r = 2m$ is required (note that the Schwarzschild horizon at $r = 2m$ is a null two - surface). They observed that the brane is a natural gravitational source for a transversable wormhole with its throat located at the horizon. This source is endowed with surface tension as an additional degree of freedom and is the place where the two copies of the exterior Schwarzschild spacetime are matched.
 
 We use in this paper the dimensionally - reduced Witten bubble spacetime \cite{EW} to obtain the Lorentzian version of the Hawking wormhole \cite{SH} which is conformal to the Rindler metric written in spherical coordinates \cite{HC1}. We find further that the stress tensor needed on the right - hand side of Einstein's equations for the Hawking wormhole to be a solution corresponds to an anisotropic fluid with negative energy density and negative radial pressure, but positive transversal pressures. Throughout the paper we shall use the geometrical units $\hbar = c = G = 1$.
  
\section{CONFORMALLY FLAT EXPANDING METRIC}

 Let us consider the Witten bubble spacetime \cite{EW}
 \begin{equation}
  ds^{2} = -g^{2} r^{2} dt^{2} + (1-\frac{4 b^{2}}{r^{2}})^{-1} d r^{2} + r^{2} \text{cosh}^{2} gt~ d\Omega^{2} + (1-\frac{4 b^{2}}{r^{2}})~ d \chi^{2}, 
  \label{1}
  \end{equation}
  with $2b \leq r \prec \infty$, $b$ and $g$ being constants, $d\Omega^{2} = d\theta^{2} + sin^{2}\theta d\phi^{2}$ being the unit two sphere and $\chi$ being the coordinate on the compactified fifth dimension. By means of Eq. (1), Witten studied the decay process (an expanding bubble) of the ground state of the original Kaluza - Klein geometry which, although stable classically, is unstable against a semiclassical barrier penetration \cite{SC}. 
  
  We are now interested in the 4 - dimensional subspace $\chi$ = const. of the Eq. (1):
 \begin{equation}
  ds^{2} = -g^{2} r^{2} dt^{2} + (1-\frac{4 b^{2}}{r^{2}})^{-1} d r^{2} + r^{2} \text{cosh}^{2} gt~ d\Omega^{2}.
 \label{2}
 \end{equation}
 Equation (2) represents the ordinary Minkowski space provided $r>>2b$, but is written in spherical Rindler coordinates \cite{HC2} (a spherical distribution of uniformly accelerated observers , with the rest - system acceleration $g$ uses this type of hyperbolically expanding coordinates). The singularity at $r = 2b$ is only a coordinate singularity, as can be seen from the isotropic form of Eq. (2), with the help of a new radial coordinate $\rho$, where 
 \begin{equation}
 r = \rho + \frac{b^{2}}{\rho}.
 \label {3}
 \end{equation}
 Therefore, the spacetime given by Eq. (2) appears now as 
  \begin{equation}
   ds^{2} = \left( 1+\frac{b^{2}}{\rho^{2}}\right)^{2} (-g^{2} \rho^{2} dt^{2} + d\rho^{2} + \rho^{2} \text{cosh}^{2} gt~ d\Omega^{2}).
 \label{4}
 \end{equation}
 We note that the coordinate transformation 
  \begin{equation}
 \bar{x} = \rho~ \text{cosh} gt~ \text{sin}\theta~ \text{cos} \phi ,~~\bar{y} = \rho~ \text{cosh} gt~ \text{sin} \theta~ \text{sin} \phi,~~\bar{z} = \rho~ \text{cosh} gt~ \text{cos}\theta, ~~\bar{t} = \rho~ \text{sinh} gt
 \label{5}
 \end{equation}
changes the previous metric into
\begin{equation}
ds^{2} = ( 1+\frac{b^{2}}{\bar{x}_{a} \bar{x}^{a}})^{2} ~\eta_{cd}~ d\bar{x}^{c}~ d\bar{x}^{d},
\label{6}
\end{equation}
which is conformally flat ($\eta_{cd} = diag (-1, +1, +1, +1))$. The latin indices run from $0$ to $3$). In addition, we have $\rho^{2} = \eta_{ab} \bar{x}^{a} \bar{x}^{b} ~\geq~0$. From now on, we take the constant $b$ to be on the order of the Planck length.

Let us observe that the geometry given by Eq. (4) becomes flat provided $\rho >>b$ (the conformal factor tends to unity ) or provided $\rho<<b$ , for which the first term in the conformal factor may be neglected. Therefore, Eq. (4) represents the Lorentzian version of the euclidean Hawking wormhole \cite{BM,SH,SW}. 

 It is interesting to treat the region $\rho~<~b$ by means of the coordinate transformation
 \begin{equation}
 \bar{\rho} = \frac{b^{2}}{\rho}.
 \label{7}
 \end{equation}
 Inserting Eq. (7) in Eq. (4), we conclude that Eq. (4) is invariant under the inversion given by Eq. (7): 
 \begin{equation}
 ds^{2} = \left( 1+\frac{b^{2}}{\bar{\rho}^{2}}\right)^{2} (-g^{2} \bar{\rho}^{2} dt^{2} + d\bar{\rho}^{2} + \bar{\rho}^{2} \text{cosh}^{2} gt ~d\Omega^{2}).
 \label{8}
 \end{equation}
 When $\rho$ varies from $b$ to infinity, $\bar{\rho} \in (0,~b]$, and the metric given by Eq. (8) becomes flat for $\bar{\rho}~<<~b$. Therefore, we may say that the $\rho~<~b$ region is the image of the $\rho~>~b$ region obtained by inversion.
 
\section{ENERGY MOMENTUM TENSOR} 

  The Hawking wormhole is well known not to be a solution to the vacuum Einstein equations. Visser \cite{MV}, by surgically grafting two Schwarzschild spacetimes together, reached the conclusion that a boundary layer (the Schwarzschild wormhole throat) contains a nonzero stress energy and separates two asymptotically flat regions. Using the junction conditions formalism, he found that the boundary layer must concentrate exotic stress energy, which violates the weak energy condition. We saw before that, according to Guendelman \textit{et al.} \cite{GKNP}, the Einstein - Rosen bridge solution does not satisfy the vacuum Einstein equations at the wormhole throat; we need a surface stress tensor on the throat. 
 
 We, instead, ask here for an energy momentum tensor on the right hand side of Einstein's equations such that the spacetime given by Eq. (4) be a solution. Using the MAPLE - GRTensor software package, we found that the nonzero components of the Einstein tensor 
 \begin{equation}
 G_{ab} \equiv R_{ab} - \frac{1}{2} g_{ab} R_{c}^{c} = \kappa T_{ab} 
 \label{9}
 \end{equation}
 are given by 
 \begin{equation}
 G_{t}^{t} = G_{\theta}^{\theta} = G_{\phi}^{\phi} = - \frac{1}{3} G_{\rho}^{\rho} = \frac{4b^{2}\rho^{4}}{(\rho^{2} + b^{2})^{4}} 
 \label{10}
 \end{equation}
 and $R_{a}^{a} = \kappa T_{a}^{a} = 0$, where $\kappa = 8 \pi$ is Einstein's constant. We see that the fluid is anisotropic and is comoving with the spherical Rindler observers ($T_{a}^{b}$ is diagonal). 
 
  By using well known notations, we have 
 \begin{equation}
 \epsilon = - T_{t}^{t} = - \frac{4b^{2}\rho^{4}}{\kappa (\rho^{2} + b^{2})^{4}}
 \label{11}
 \end{equation}
 for the energy density and 
 \begin{equation}
 T_{\rho}^{\rho} = p_{\rho} = -3 p_{\theta} = -3 p_{\phi} = \frac{-12 b^{2} \rho^{4}}{\kappa (\rho^{2} + b^{2})^{4}}
 \label{12}
 \end{equation}
 for the radial and the transverse pressures, respectively. Even though the fluid is anisotropic, with pressures not only of different values but also of different signs, we can define a mean value
 \begin{equation}
 p_{m} = \frac{p_{\rho} + p_{\theta} + p_{\phi}}{3} = \frac{\epsilon}{3},
 \label{13}
 \end{equation}
 as for a null fluid.
 
 We shall consider the energy to be mostly located on the throat of the wormhole given by using Eq. (4), which is in accordance with a negative energy density ($\epsilon < 0$) on the throat. From Eq. (11), one infers that the case $\rho >> b$ leads to an expression for $\epsilon$ that does not depend on the Newton constant $G$:
 \begin{equation}
 \epsilon = - \frac{4b^{2}}{8 \pi G \rho^{4}} = - \frac{\hbar c}{2\pi \rho^{4}}, 
 \label{14}
 \end{equation}
 where $b$ has been replaced with the Planck length $\sqrt{G \hbar/c^{3}}$. In other words, $\epsilon$ has a purely quantum origin in this region, and is similar to the Casimir energy density between two perfectly conducting parallel plates.\\
    
\section{NULL GEODESIC EXPANSION}

  According to Hochberg and Visser \cite{HV} a natural local geometric characterization of the existence and the location of a wormhole throat is possible. They argued that the throat is an extremal hypersurface of minimal area ( minimal anti-trapped surface) with respect to deformations in the null direction $\lambda$  - the parameter on the null geodesic (the tangent vector is $k^{a} = dx^{a}/d\lambda$). In addition, the two - dimensional spatial hypersurface to which the null geodesic congruence is orthogonal will be the wormhole throat provided the scalar expansion $\Theta = 0$
 and $d\Theta/d\lambda > 0$ hold on that surface (the flare - out condition generalized for a dynamic wormhole). 
 
 According to the previous authors' prescription, we take the null geodesic congruence orthogonal to a compact two - dimensional surface embedded in the spacetime. We introduce a future-directed outgoing null 4 - vector $k_{+}^{a}$ , a future directed ingoing null 4 - vector $k_{-}^{a}$ and a spatial orthogonal projection tensor $\gamma^{ab}$ which obey the following relations:
 \begin{equation}
 \begin{split}
 k_{+}^{a} k_{+a} = k_{-}^{a} k_{-a} =  0,~~~~ k_{+}^{a} k_{-a} = k_{-}^{a} k_{+a} = -1,\\
 k_{\pm}^{a} \gamma_{ab} = 0,~~~~\gamma^{a}_{b} \gamma^{bc} = \gamma^{ac}.
 \label{15}
 \end{split}
 \end{equation}
 The full spacetime metric may be decomposed into terms involving the previous vectors (see also Ref. 14): 
 \begin{equation}
 g_{ab} = \gamma_{ab} - k_{-a} k_{+b} - k_{+a} k_{-b}. 
 \label{16}
 \end{equation}
 We note that the spacelike hypersurface must have two sides and that $+$ and $-$ are just two labels for the two null direction. Defining the purely spatial tensors $v_{ab}^{\pm} = \gamma_{a}^{c} \gamma_{b}^{d} \nabla_{c} k_{\pm d}$, we have \cite{HV} 
 \begin{equation}
 \Theta_{\pm} = \gamma^{ab} v_{ab}^{\pm} = \gamma^{ab} \nabla _{a} k_{\pm b},
 \label{17}
 \end{equation}
 where the expansions $\Theta_{\pm}$ represents the trace of $v_{ab}^{\pm}$ and $\gamma_{ab} \gamma^{ab} = 2$.
 
 The Raychaudhuri equation for the congruence of null geodesics can be written as 
 \begin{equation}
 \dot{\Theta_{\pm}} + 2 (\sigma^{2}_{\pm} - \Omega^{2}_{\pm}) + \frac{1}{2} \Theta^{2}_{\pm} = -R_{ab} k^{a}_{\pm} k^{b}_{\pm}, 
 \label{18}
 \end{equation}
 where $\dot{\Theta} = d\Theta/d\lambda = k^{a}\nabla_{a} \Theta,~2\sigma^{2}_{\pm} = \sigma^{\pm ab} \sigma_{\pm ab}, ~2\Omega^{2}_{\pm} = \Omega^{\pm ab} \Omega_{\pm ab}$ and the shear tensor $\sigma_{ab} $ and the vorticity tensor $\Omega_{ab}$ are given, respectively, by
 \begin{equation}
 \sigma_{ab}^{\pm} = \frac{1}{2} (v_{ab}^{\pm} + v_{ba}^{\pm}) - \frac{1}{2} \Theta^{\pm} \gamma_{ab}
 \label{19}
 \end{equation}
 and
 \begin{equation}
 \Omega_{ab}^{\pm} = \frac{1}{2} (v_{ab}^{\pm} - v_{ba}^{\pm}).
 \label{20}
 \end{equation}
 Let us find now the expansion $\Theta_{\pm}$ for our wormhole given by Eq. (4) and, from here, the location of the throat, namely, the surface where $\Theta_{\pm} = 0$. By use of Eqs. (15) and (16), it is an easy task to obtain 
 \begin{equation}
 \begin{split}
 k^{+a} = ( \frac{1}{\omega^{2}}, - \frac{g \rho}{\omega^{2}}, 0, 0 ),~~~k^{-a} = ( \frac{1}{2g^{2}\rho^{2}}, \frac{1}{2g \rho}, 0, 0 ),\\ 
 \gamma_{ab} = (0, 0, \omega^{2}\rho^{2} \text{cosh}^{2} gt, \omega^{2}\rho^{2} \text{cosh}^{2}gt~ \text{sin}^{2}\theta ),
 \end{split}
 \label{21}
 \end{equation}
 with $\omega = 1 + (b^{2}/\rho^{2})$ and the components of the 4 - vectors being in order, $t, \rho, \theta,$ and $\phi$.
 
 With the help of the following expressions for the Christoffel symbols,
 \begin{equation}
 \begin{split}
 \Gamma_{\theta \theta}^{t} = \frac{\Gamma_{\phi \phi}^{t}}{\text{sin}^{2}\theta} = \frac{1}{g} \text{cosh}gt~ \text{sinh}gt,~~~\Gamma_{\theta \theta}^{\rho} = - \frac{\rho (\rho^{2} - b^{2}) \text{cosh}^{2} gt}{\rho^{2} + b^{2}}= \\
 = \frac{\Gamma_{\phi \phi}^{\rho}}{\text{sin}^{2}\theta},~~~\Gamma_{\rho t}^{t} = \frac{\rho^{2} - b^{2}}{\rho (\rho^{2} + b^{2})},~~~\Gamma_{\rho \rho}^{\rho} = - \frac{2b^{2}}{\rho (\rho^{2} + b^{2})}, 
 \end{split}
 \label{22}
 \end{equation}
 we get 
 \begin{equation}
 \Theta_{+} = \frac{2g}{\omega^{3} \text{cosh}gt} \left[(1+\frac{b^{2}}{\rho^{2}}) \text{sinh}gt - (1-\frac{b^{2}}{\rho^{2}}) \text{cosh}gt\right]
 \label{23}
 \end{equation}
 and
 \begin{equation}
 \Theta_{-} = \frac{1}{g\rho^{2} \omega \text{cosh}gt} \left[(1+\frac{b^{2}}{\rho^{2}}) \text{sinh}gt + (1 - \frac{b^{2}}{\rho^{2}})\text{cosh}gt\right].
 \label{24}
 \end{equation}
 The conditions $\Theta_{\pm} = 0$ for the throats give us 
 \begin{equation}
 \rho_{+} (t) = b~ e^{gt} ,~~~~\rho_{-} (t) = b ~e^{-gt} ,
 \label{25}
 \end{equation}
 with $\rho_{\pm}(0) = b$. A time reversal changes $\rho_{+}(t)$ to $\rho_{-}(t)$. We see that while one throat is expanding, the other is shrinking. Since the spacetime in Eq. (4) is invariant under an inversion $\rho \rightarrow \bar{\rho} = b^{2}/\rho$, we can replace $\rho_{+}(t) < b ~(t < 0)$ with $\bar{\rho}_{+}(t) > b~ (t < 0)$, obtaining in this way two copies of the region $[b, \infty): \bar{\rho}_{+} (t)$ for $t<0$ and $\rho_{+} (t)$ for $t>0$. Therefore, one throat comes from infinity at $t \rightarrow - \infty$ , reaches $\bar{\rho_{+}} (t) = b$ at $t = 0$ and returns to infinity at $t \rightarrow \infty$. Similar arguments may be applied for the domain $[0,b]$. 
 
 An inspection of the null radial geodesics in the geometry given by Eq. (4) yields
 \begin{equation}
 - g^{2} \rho^{2} dt^{2} + d\rho^{2} = 0.
 \label{26}
 \end{equation}
 Therefore, $\rho_{\pm} (t) = \rho_{0}~ e^{\pm gt}$, with $\rho_{0} = \rho (0)$. The conformal factor does not change, of course, the null trajectories, so the previous null curves correspond to those viewed by a spherical distribution of uniformly accelerated observers in Minkowski space with acceleration $g$. In addition, we notice that the particular null radial geodesic with $\rho(0) = b$ represents the expanding (contracting) wormhole throat found before. Reversing the logic, we conjecture that the propagation of light follows the throat of a pre - existing wormhole. The expanding throat corresponds to retarded radiation ($t > 0$) and the contracting throat corresponds to the advanced radiation ($t < 0$). In other words, our empty space is full of expanding - contracting throats that drag the light with them. The throats correspond, in the conformally flat metric given by Eq. (6), to the light cones $\bar{r} = \pm{\bar{t}}$ or $\bar{r} = \pm {\bar{t}} + b$, with $\bar{r} = \sqrt{\bar{x}^{2} + \bar{y}^{2} + \bar{z}^{2}}$. On the grounds of the previous results, we conclude that the geometry given by Eq. (6) is more suitable than the Minkowski spacetime for the spacetime felt by an inertial observer. We have a nonvanishing stress tensor near the wormhole throat (light cone), but the components of $T_{ab}$ tends to zero far from the light cone. We notice that any geodesic with $\rho_{0} \neq b$ does not represents a wormhole throat. That is a consequence of the fact that the spacetime given by Eq. (4) is not invariant under $\rho$ - translations.
 
 Let us check now whether the minimal conditions $\dot{\Theta}_{\pm} \equiv d\Theta_{\pm}/d\lambda \geq 0$ are fulfilled. We have 
 \begin{equation}
 k_{+}^{a} \nabla_{a} \Theta_{+} \equiv \frac{d\Theta_{+}}{d\lambda}|_{\Theta_{+} = 0} = \frac{16 b^{2} g^{2}}{\rho^{2} \omega^{6}},~~~~k_{-}^{a}\nabla_{a} \Theta_{-} \equiv  \frac{d\Theta_{-}}{d\lambda}|_{\Theta_{-} = 0} = \frac{4b^{2}}{g^{2} \rho^{6} \omega^{2}}.
 \label{27}
 \end{equation}
 Hence, both $\dot{\Theta}_{\pm}$ are everywhere positive, and the flare - out condition is satisfied. That is in accordance with the negative value of the energy density $\epsilon$ (exotic matter).
 
 It is worth noting that we use a non - affine connection for our congruence of null geodesics. Therefore, the Raychaudhuri equation (18) must have an extra term, $- K \Theta_{\pm}$, on its left hand side, where $K$ is obtained from \cite {KS, EFO}
 \begin{equation}
 a^{b} \equiv k_{\pm}^{a} \nabla_{a} k_{\pm}^{b} = K_{\pm} k_{\pm}^{b}.
 \label{28}
 \end{equation}
 This is the geodesic equation in a non-affine parameterization, and $K$ plays the role of surface gravity \cite{RW}. From Eq. (21) for $k_{\pm}^{a}$,  we find the components of the accelerations $a_{\pm}^{b}$ :
 \begin{equation}
  a_{+}^{b} = ( \frac{-2g}{\omega^{4}}, \frac{2g^{2}\rho}{\omega^{4}}, 0, 0),~~~~a_{-}^{b} = (-\frac{b^{2}}{g^{3} \rho^{6} \omega}, -\frac{b^{2}}{g^{2} \rho^{5} \omega}, 0, 0) ,
  \label{29}
  \end{equation}
  with $a_{+}^{b}a_{b}^{+} = a_{-}^{b}a_{b}^{-} = 0$.    Therefore, Eqs. (27) and (29) yield
  \begin{equation}
  K_{+} = - \frac{2g}{\omega^{2}} ,~~~~K_{-} = - \frac{2b^{2}}{g\rho^{4} \omega}.
  \label{30}
  \end{equation}
  Using the above equations for $K_{\pm}$, keeping in mind that $\sigma_{ab}^{\pm}$ and $\Omega_{ab}^{\pm}$ vanish in our spacetime, Eq. (4), and  that, with the help of Eqs. (11) and (12), the right hand side of Raychaudhuri's equations are given by 
 \begin{equation}
 \kappa T_{ab} k_{+}^{a} k_{+}^{b} = - \frac{16 b^{2} g^{2}}{\rho^{2} \omega^{6}},~~~~\kappa T_{ab} k_{-}^{a} k_{-}^{b} = - \frac{4b^{2}}{g^{2} \rho^{6} \omega^{2}},
 \label{31}
 \end{equation}
  we convince ourselves that the ($\pm$) Raychaudhuri equations  are satisfied.
  
\section{ANISOTROPIC STRESS TENSOR}

  Let us show now that the energy momentum tensor $T_{ab}$ with the components given by Eq. (11) - (12) can be expressed in the general form characterizing an anisotropic fluid \cite{HMO}:
 \begin{equation}
 T_{ab} = ( \epsilon + p_{\bot}) k_{a} k_{b} + p_{\bot} g_{ab} + ( p_{\rho} - p_{\bot}) s_{a} s_{b}, 
 \label{32}
 \end{equation}
 where $p_{\bot}$ is the transversal pressure and $s_{a}$ is a spacelike vector in the direction of anisotropy. Taking $p_{\bot} = p_{\theta} = p_{\phi} $, we get $\epsilon + p_{\bot} = 0$, and the first term on the right hand side of Eq. (32) is vanishing. With $s_{a} = ( 0, \omega, 0, 0 ) $ and $s_{a} s^{a} = 1$, our stress tensor acquires the simple form
 \begin{equation}
 T_{ab} = ( \epsilon + p_{\rho} ) s_{a} s_{b} - \epsilon g_{ab}, 
 \label{33}
 \end{equation}
 which corresponds to an anisotropic fluid due to the fact that the radial and the transversal pressures are different. 
 
 Let us note that the energy density on the throats, $\rho_{\pm} (t) = b ~e^{\pm gt}$, can be written as 
 \begin{equation}
 \epsilon (t) = - \frac{1}{\kappa \rho_{th}^{2} (t)},
 \label{34}
 \end{equation}
 where $\rho_{th} (t) = 2b~ \text{cosh}^{2} gt$ is the time - dependent radius of the throats obtained from the metric
 \begin{equation}
 ds^{2}_{th} = 4b^{2} \text{cosh}^{2} gt~ (d\theta^{2} + \text{sin}^{2} \theta d\phi^{2} ).
 \label{35}
 \end{equation}
 It is invariant under time reversal and the inversion $\rho \rightarrow b^{2} /\rho$.\\
 
\section{CONCLUSIONS}

  We introduced in this paper an energy momentum tensor on the right hand side of Einstein's equations, which corresponds to an anisotropic fluid with negative energy density $\epsilon = - p_{\theta} = - p_{\phi}$ in order for the Lorentzian Hawking wormhole to be a solution. For a radial coordinate $\rho$ much greater than the Planck length ($\rho >> b$), we found the energy density $\epsilon$ no longer depended on Newton's constant $G$, and that $\epsilon$ had a purely quantum origin. The wormhole geometry is invariant under the spatial inversion $\rho \rightarrow \bar{\rho} = b^{2}/\rho$. Therefore, the two regions $\rho << b$ and $\rho >> b$, are equivalent. We proved also that the scalar expansions $\Theta_{\pm}$ of the outgoing and the ingoing congruence of null geodesics are vanishing for $\rho_{\pm}(t) = b~e^{\pm gt}$ , which suggests that the throats of the dynamic wormholes expand (shrink) exponentially, $\rho(t)$ being identified with the null radial geodesics. 
 
 To check the Raychaudhuri equations, we kept in mind that we had not used an affine parameterization for the null congruence, having added an extra term proportional to the scalar expansion. The stress tensor needed to satisfy Einstein's equations for the spacetime given by Eq. (4), located near the throat (or near the light cone in Cartesian coordinates), is traceless, the null energy condition is violated and the flare - out condition is obeyed.

\end{document}